\documentclass[%
 reprint,
amsmath,amssymb,
aps,
pra,
floatfix,
]{revtex4-1}

\usepackage[T1]{fontenc}
\usepackage[utf8]{inputenc}
\usepackage[english]{babel}
\usepackage{dsfont}

\addto\captionsenglish{}

\usepackage{subfigure}

\usepackage{comment}

\usepackage[colorlinks]{hyperref}
\hypersetup{
	colorlinks = true,
	linkcolor = red,
	citecolor = {green},
	urlcolor = {blue}
}

\usepackage{float}

\usepackage{graphicx}
\usepackage{dcolumn}
\usepackage{bm}

\begin{document}

\title{Quantum fluctuations beyond the Gutzwiller approximation \texorpdfstring{\\ in the Bose-Hubbard model}{in the Bose-Hubbard model}}

\author{Fabio Caleffi$^{1,2}$}
\email{fcaleffi@sissa.it}
\author{Massimo Capone$^{1,3}$}
\author{Chiara Menotti$^{2}$}
\author{Iacopo Carusotto$^{2,4}$}
\author{Alessio Recati$^{2,4}$}
\affiliation{$^1$International School for Advanced Studies (SISSA), Via Bonomea 265, I-34136 Trieste, Italy}
\affiliation{$^2$INO-CNR  BEC  Center  and  Dipartimento di  Fisica,  Universit\`a di Trento,  38123  Povo, Italy}
\affiliation{$^3$CNR-IOM Democritos, Via Bonomea 265, I-34136 Trieste, Italy}
\affiliation{$^4$Trento Institute for Fundamental Physics and Applications, INFN, 38123, Trento, Italy}
\email{alessio.recati@cnr.it}

\date{\today}

\begin{abstract}
We develop a quantum many-body theory of the Bose-Hubbard model based on the canonical quantization of the action derived from a Gutzwiller mean-field ansatz.  
Our theory is a systematic generalization of the Bogoliubov theory of weakly-interacting gases. The control parameter of the theory, defined as the zero point fluctuations on top of the Gutzwiller mean-field state, remains small in all regimes. The approach provides accurate results throughout the whole phase diagram, from the weakly to the strongly interacting superfluid and into the Mott insulating phase.
As specific examples of application, we study the two-point correlation functions, the superfluid stiffness, the density fluctuations, for which quantitative agreement with available quantum Monte Carlo data is found. In particular, 
the two different universality classes of the superfluid-insulator quantum phase transition at integer and non-integer filling are recovered.
\end{abstract}

\maketitle

\section{Introduction}
The Hubbard model is one of the most celebrated models of quantum condensed matter theory. The main reason is probably the widespread belief that its two-dimensional fermionic version holds the key to understand how high-temperature superconductivity emerges upon doping a Mott insulator \cite{Anderson1987, Lee2006}. Its central feature is the competition between the kinetic energy term, which favors delocalized states, and the local Coulomb repulsion which favors localization \cite{Castellani1979, Vollhardt1984}.
In the two-dimensional fermionic model this physics is however somewhat hidden by the presence of other phases bridging between the Mott insulator, the superconducting and the metallic states, including the celebrated pseudogap \cite{Gull2013} and charge-ordered \cite{Imada2018} phases. 

The archetypal competition between the kinetic and interaction energies is found in the bosonic version of the model, the so-called Bose-Hubbard (BH) model~\cite{fisher}, where it manifests itself as a direct quantum phase transition between a superfluid and a Mott insulator. As a consequence of its paradigmatic nature, this transition has attracted an enormous experimental interest in the last years, fostering implementations with cold atoms trapped in optical lattices \cite{jaksch_bruder_cirac_gardiner_zoller,greiner_bloch, lattice_depth_modulation_1, folling_bloch, bakr_greiner, bragg, lattice_depth_modulation_2, pitaevskii_stringari} and, more recently, implementations with photons in the novel context of non-equilibrium phase transitions \cite{carusotto2013quantum, ma2019dissipatively,carusotto2019}. 

On the theoretical side, a very common approach to the BH model is based on the Gutzwiller ansatz.
While many important features of both the superfluid and the insulating phases are accurately captured by the Gutzwiller wave function, its local, site-factorized form typically makes physical quantities involving off-site quantum correlations to be missed.
In the weakly-interacting regime, a Bogoliubov approach for the fluctuations around the mean-field Gross-Pitaevskii (GP) ground state of a dilute Bose-Einstein condensate provides an accurate description of the equilibrium state and of the excitations of the gas~\cite{castin, pitaevskii_stringari}, including quantum correlations between particles~\cite{Castin-IC2003}. 
In the strongly-interacting regime, however, the GP mean-field theory and the Bogoliubov approach based on it become clearly inadequate. The rich physics of the strongly interacting BH model across the Mott-superfluid transition and specifically in the insulating phase has been attacked through a number of different approaches, ranging from semi-analytical methods as RPA \cite{sheshandri_krishnamurthy_pandit_ramakrishnan, oosten_straten_stoof, menotti_trivedi}, slave boson representation \cite{slave_boson, roscilde}, time-dependent Gutzwiller approximation \cite{krutitsky_navez, BEC}, to numerical techniques including quantum Monte Carlo methods \cite{qmc_3, qmc_2, qmc_1, qmc_6}, bosonic Dynamical Mean-Field Theory (B-DMFT) \cite{bdmft_vollhardt, bdmft_tong, bdmft_pollet} and Numerical Renormalization Group (NRG) \cite{dupuis_1, dupuis_2}. All these methods provide qualitatively concordant results on the phase diagram as well as on the spectral properties of the model. The collective phonon excitations of the Bogoliubov theory of dilute condensates are replaced by a multi-branch spectrum of excitations~\cite{ohashi_kitaura_matsumoto, menotti_trivedi,BEC,krutitsky_navez}, containing in particular the gapless Goldstone mode and a gapped (also refereed to as Higgs) mode on the superfluid side and the particle/hole excitations in the insulating phase (see e.g.~\cite{sachdev,huber_altman_buchler_blatter,BEC}).

In spite of these remarkable advances, a complete, easily tractable and physically intuitive description of the collective excitations and their fingerprint on quantum observables across the whole phase diagram of the model is still lacking. 
In particular, the development of non-local correlations across the Mott-superfluid transition and the proper characterization of the strongly-interacting superfluid state and of its excitation modes remain a challenging problem.

In this paper, we combine the successful features of the Gutzwiller and Bogoliubov approaches to develop a new strategy to systematically quantize the time-dependent Gutzwiller ansatz. In spite of the local nature of the Gutzwiller ansatz -- see Eq. \eqref{ansatz} below --, the accurate description of the excitations and, in particular, of their zero-point fluctuations allows to correctly reproduce the non-local many-body correlations in the different phases, as well as the different critical behaviours of the commensurate and incommensurate phase transitions~\cite{fisher, sachdev}.
Time-dependent Gutzwiller approaches addressing the linear-response dynamics in the BH model~\cite{krutitsky_navez} and in lattice Fermi systems~\cite{fabrizio} have been recently developed. The advantage of our formalism is that it directly includes quantum fluctuations of the collective modes and could naturally incorporate those effects beyond linearised fluctuations that stem from interactions between quasi-particles. This is essential to successfully tackle problems such as the finite quasi-particles lifetime via Beliaev-like nonlinear interaction processes and the quantum correlations between the products of their decay, which will be the subject of future investigations.

The paper is organized as follows. Sect.~\ref{section_model} is devoted to the derivation of the quantum Gutzwiller theory for the Bose-Hubbard model. The original features of the method are highlighted and its advantages and disadvantages are discussed in comparison to other approaches. In Sect.~\ref{section_results}, we present the predictions of the quantum Gutzwiller method for observables in which local and non-local quantum correlations strongly modify the standard mean-field picture, such as two-point correlation functions, superfluid density and pair correlations. We conclude in Sect.~\ref{section_conclusions} with an outlook on future studies and on possible extensions of the quantum formalism introduced in this work.

\section{Model and theory}\label{section_model}

In Subsect.~\ref{subsec:basics}, we briefly review the basic concepts of the Bose-Hubbard model and of the $\mathbb{C}$-number Gutzwiller ansatz needed to develop the quantum Gutzwiller approach presented in the following Subsects.~\ref{subsec:quantumGutz} to \ref{protocol}. The quantum Gutzwiller method is put into perspective and compared with other approaches in Subsect.~\ref{subsec:perspective}.

\subsection{Lagrangian formulation within the Gutzwiller ansatz}
\label{subsec:basics}

We consider the three-dimensional BH model

\begin{equation}\label{hamiltonian}
\small
\hat{H} = -J \sum_{\langle \mathbf{r}, \mathbf{s} \rangle} \left( \hat{a}^{\dagger}_{\mathbf{r}} \, \hat{a}_{\mathbf{s}} + \text{h.c.} \right) + \frac{U}{2} \sum_{\mathbf{r}} \hat{n}_{\mathbf{r}} \left( \hat{n}_{\mathbf{r}} - 1 \right) - \mu \sum_{\mathbf{r}} \hat{n}_{\mathbf{r}}
\end{equation}
where $J$ is the hopping amplitude, $U$ the on-site interaction, $\mu$ the chemical potential, while $\langle \mathbf{r}, \mathbf{s} \rangle$ labels all pairs of nearest-neighboring sites. The annihilation and creation operators of a bosonic particle at site $\mathbf{r}$ are $\hat{a}_{\mathbf{r}}$ and $\hat{a}^{\dagger}_{\mathbf{r}}$ respectively, and $\hat{n}_{\mathbf{r}} = \hat{a}^{\dagger}_{\mathbf{r}} \, \hat{a}_{\mathbf{r}}$ is the local density operator.

We use the Gutzwiller ansatz ~\cite{Krauth1992,krutitsky, krutitsky_navez}

\begin{equation}\label{ansatz}
\small
| \Psi_G \rangle = \bigotimes_{\mathbf{r}} \sum_n c_n{\left( \mathbf{r} \right)} \, | n, \mathbf{r} \rangle,
\end{equation}
where the wave function is site-factorized
and the complex amplitudes $c_n{\left( \mathbf{r} \right)}$ are variational parameters with normalization condition $\sum_n \left| c_n{\left( \mathbf{r} \right)} \right|^2 = 1$ to reformulate the Bose-Hubbard model in terms of the following Lagrangian functional 
\begin{eqnarray}\label{lagrangian}
\mathfrak{L}{\left[ c, c^{*} \right]} &=& \Big\langle \Psi_G \Big| \, i \, \hbar \, \partial_t - \hat{H} \, \Big| \Psi_G \Big\rangle = \\
&=& \frac{i \, \hbar}{2} \sum_{\mathbf{r},n}  [c^{*}_n(\mathbf{r}) \dot{c}_n(\mathbf{r}) - \textrm{c.c.}] \nonumber \\
&+ J& \sum_{\langle \mathbf{r}, \mathbf{s} \rangle} \left[ \psi^{*}{\left( \mathbf{r} \right)} \, \psi{\left( \mathbf{s} \right)} + \text{c.c.} \right] - \sum_{\mathbf{r},n} H_n \left| c_n{\left( \mathbf{r} \right)} \right|^2. \nonumber
\end{eqnarray}
In the previous equation, the dot indicates the temporal derivative,
\begin{equation}
H_n = \frac{U}{2}  n \left( n - 1 \right) - \mu \, n
\end{equation}
are the on-site terms in the Hamiltonian and 
\begin{equation}
\psi{\left( \mathbf{r} \right)}=
\big\langle \hat{a}_{\mathbf{r}} \big\rangle = \sum_n \sqrt{n} \, c^{*}_{n - 1}{\left( \mathbf{r} \right)} \, c_n{\left( \mathbf{r} \right)}
\end{equation}
is the mean-field order parameter. 
In this formulation, the conjugate momenta of the parameters ${c}_n{\left( \mathbf{r} \right)}$ are ${c}_n^{*}{\left( \mathbf{r} \right)} =\partial \mathfrak{L} / \partial {\dot{c}_n}{\left( \mathbf{r} \right)}$.
The classical Euler-Lagrange equations associated to Lagrangian (\ref{lagrangian}) are the {\it time-dependent Gutzwiller equations} as derived, e.g., in \cite{sheshandri_krishnamurthy_pandit_ramakrishnan, krutitsky_navez} and from which the excitation spectrum can be determined.
In the uniform system the stationary solutions are homogeneous. The system is in a Mott Insulator (MI) state if $\psi{\left( \mathbf{r} \right)}=0$ and in a Superfluid (SF) state otherwise.  

\begin{figure*}[!htp]
    \centering
	\includegraphics[width=1.0\linewidth]{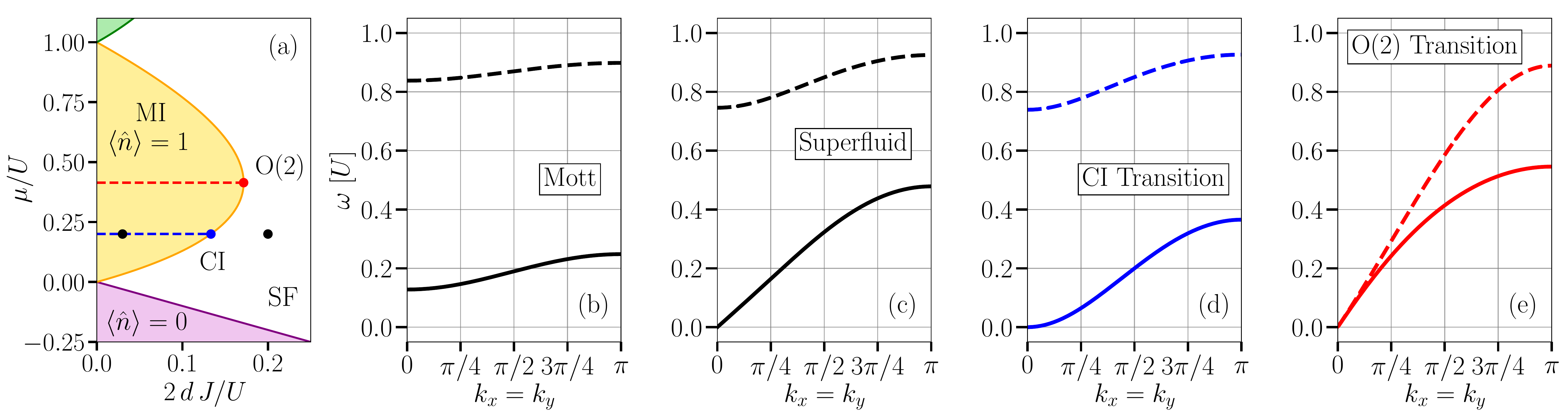}
    \caption{Panel (a): mean-field Gutzwiller phase diagram around the $\langle \hat{n} \rangle = 1$ Mott lobe. The black points refer to the MI and SF spectra shown in panels (b, c), while the blue and red points indicate the CI and $\text{O}{\left( 2 \right)}$ critical points respectively.
    Panel (b): energy spectra of hole (solid line) and particle (dashed line) excitations in the MI phase for $\mu/U = 0.2$ and $2 \, d \, J/U = 0.03$.
    Panel (c): Goldstone mode (solid line) and Higgs mode (dashed line) energy dispersion in the strongly-correlated SF phase for $\mu/U = 0.2$ and $2 \, d \, J/U = 0.2$.
    Panel (d): excitation spectrum at the CI critical point corresponding to $\mu/U = 0.2$ and $2 \, d \, J/U = 0.1\overline{3}$, see blue dot in panel (a). 
    Panel (e): excitation spectrum at the $\text{O}{\left( 2 \right)}$-invariant critical point, see red dot in panel (a). Both modes become gapless and present a linear dispersion.}
    \label{fig:figure1}
\end{figure*}

The spectrum of the collective modes $\omega_{\alpha,\mathbf{k}}$ is plotted in \autoref{fig:figure1} in different regions of the phase diagram shown in panel (a). In the MI phase [panel (b)], the two lowest excitation branches are the gapped particle and hole excitations. In the SF phase [panel (c)], the lowest of them becomes the gapless Goldstone mode of the broken U(1) symmetry. The other gapped excitation is often referred to as the Higgs mode~\cite{pekker_varma, podolsky_sachdev, podolsky_higgs, BEC} and is related to the fluctuations of the amplitude of the order parameter in some specific region of the phase diagram~\cite{BEC}.

The quantum phase transition from the MI to the SF phase can belong to two different universality classes \cite{fisher, sachdev} depending on whether the transition is crossed while changing the density -- the so-called commensurate-incommensurate (CI) transition [blue point in panel (a)]  -- or it is crossed at a fixed and commensurate filling (at the tip) -- the so-called $\text{O}{\left( 2 \right)}$ transition [red point in panel (a)]. At the CI transition points only one mode becomes gapless (the Goldstone branch in the SF), whereas the other mode is gapped and related to the particle or the hole branch of the MI depending on the chemical potential [panel (d)]. On the other hand, at the tip critical point both modes become gapless [panel (e)].

\subsection{The quantum Gutzwiller theory}
\label{subsec:quantumGutz}
In order to go beyond the Gutzwiller ansatz reviewed in the previous subsection, it is natural to consider how quantum (and thermal) effects populate the excitation modes of the system and to address how they affect the observable quantities.    
We include quantum fluctuations by building a theory of the excitations starting from Lagrangian~(\ref{lagrangian}) via canonical quantization~\cite{cohen, blaizot_ripka}, namely promoting the coordinates and their conjugate momenta to operators and imposing equal-time canonical commutation relations

\begin{equation}
\left[{\hat{c}_n}{\left( \mathbf{r} \right)}, {\hat{c}_m}^{\dagger}{\left( \mathbf{s} \right)} \right] = \delta_{\mathbf{r},\mathbf{s}}\,\delta_{n,m}.
\end{equation}
In analogy with the Bogoliubov approximation for dilute Bose-Einstein condensates~\cite{pitaevskii_stringari,castin}, we expand the operators ${\hat c}_n$ around their ground state values $c_n^0$, obtained by minimizing the energy $\langle \Psi_G | \hat{H}|\Psi_G\rangle$, as 

\begin{equation}\label{c_operator}
\hat{c}_n{\left( \mathbf{r} \right)} = \hat{A}({\mathbf r}) \, c^0_n + \delta \hat{c}_n{\left( \mathbf{r} \right)}.
\end{equation}
The {\it normalization operator} ${\hat A}({\mathbf r})$ is a function of $\delta \hat{c}_n\left( \mathbf{r} \right)$ and $\delta \hat{c}^\dagger_n\left( \mathbf{r} \right)$ and ensures the proper normalization $\sum_n \hat{c}^{\dagger}_n{\left( \mathbf{r} \right)} \, \hat{c}_n{\left( \mathbf{r}\right)} = \hat{\mathds{1}}$. 
By restricting to local fluctuations orthogonal to the ground state $\sum_n  \delta \hat{c}^\dagger_n{\left( \mathbf{r} \right)} \, c^{0}_n  = 0$ one has
\begin{equation}
\hat{A}({\mathbf{r}}) = \left[ 1 - \sum_n \delta \hat{c}^{\dagger}_n{\left( \mathbf{r} \right)} \, \delta \hat{c}_n{\left( \mathbf{r}\right)} \right]^{1/2}.
\end{equation}

In a homogeneous system, it is convenient to work in momentum space by writing
\begin{equation}\label{FT}
\delta \hat{c}_n{\left( \mathbf{r} \right)} \equiv N^{-1/2} \sum_{\mathbf{k} \in \text{BZ}} e^{i \mathbf{k} \cdot \mathbf{r}} \, \delta \hat{C}_n{\left( \mathbf{k} \right)}.
\end{equation}
Inserting Eq.~(\ref{FT}) in $\langle \Psi_G | \hat{H}|\Psi_G\rangle$ and keeping only terms up to the quadratic order in the fluctuations, we obtain 
\begin{equation}\label{H_2}
\small
\hat{H}^{\left( 2 \right)} = E_0 + \frac{1}{2} \sum_{\mathbf{k}}
[\delta \underline{\hat{C}}^{\dagger}{\left( \mathbf{k} \right)} , -\delta \underline{\hat{C}}{\left( -\mathbf{k} \right)}]
\,\hat{\mathcal{L}}_{\mathbf{k}}
\begin{bmatrix}
\delta \underline{\hat{C}}{\left( \mathbf{k} \right)} \\
\delta \underline{\hat{C}}^{\dagger}{\left( -\mathbf{k} \right)}
\end{bmatrix}\,,
\end{equation}
where $E_0$ is the mean-field ground state energy, the vector $\delta \underline{\hat{C}}(\mathbf{k})$ contains the components $\delta{\hat{C}}_n(\mathbf{k})$, and $\hat{\mathcal{L}}_{\mathbf{k}}$ is a pseudo-Hermitian matrix, the explicit expression of which is given in App.~\ref{app_L_k}.
A suitable Bogoliubov rotation of the Gutzwiller operators %
\begin{equation}\label{rotation}
\delta \hat{C}_n{\left( \mathbf{k} \right)} = \sum_{\alpha} u_{\alpha, \mathbf{k}, n} \, \hat{b}_{\alpha, \mathbf{k}} + \sum_{\alpha} v^{*}_{\alpha, -\mathbf{k}, n} \, \hat{b}^{\dagger}_{\alpha, -\mathbf{k}}
\end{equation}
recasts the quadratic form \eqref{H_2} into a diagonal form
\begin{equation}\label{H_2_diag}
\hat{H}^{\left( 2 \right)} = \hbar \sum_{\alpha} \sum_{\mathbf{k}} \omega_{\alpha, \mathbf{k}} \, \hat{b}^{\dagger}_{\alpha, \mathbf{k}} \hat{b}_{\alpha, \mathbf{k}},
\end{equation}
where each $\hat{b}^{\dagger}_{\alpha, \mathbf{k}}$ corresponds to a different many-body excitation mode with frequencies $\omega_{\alpha,\mathbf{k}}$, labeled by its momentum $\mathbf{k}$ and branch index $\alpha$. Bosonic commutation relations between the annihilation and creation operators $\hat{b}_{\alpha, \mathbf{k}}$ and  $\hat{b}^\dagger_{\alpha, \mathbf{k}}$ 
\begin{equation}[\hat{b}_{\alpha, \mathbf{k}},\hat{b}^{\dagger}_{\alpha', \mathbf{k'}}]=\delta_{\mathbf{k},\mathbf{k}'}\,\delta_{\alpha,\alpha'}
\end{equation}
are enforced by choosing the usual Bogoliubov normalization condition 
\begin{equation}
\underline{u}^{*}_{\alpha, \mathbf{k}} \cdot \underline{u}_{\beta, \mathbf{k}} - \underline{v}^{*}_{\alpha, -{\mathbf{k}}} \cdot \underline{v}_{\beta, -\mathbf{k}} = \delta_{\alpha \beta}\,,\end{equation}
where the vectors $\underline{u}_{\alpha, \mathbf{k}}$ ($\underline{v}_{\alpha, \mathbf{k}}$) contain the components $u_{\alpha, \mathbf{k}, n}$ ($v_{\alpha, \mathbf{k}, n}$). As a direct consequence of the spectral properties of $\mathcal{L}_{\mathbf{k}}$, the fluctuation operators $\delta \hat{c}_n{\left( \mathbf{r} \right)}$ satisfy the quasi-bosonic commutation relations
\begin{equation}\label{commutations}
\left[{\delta \hat{c}_n}{\left( \mathbf{r} \right)}, {\delta \hat{c}_m}^{\dagger}{\left( \mathbf{s} \right)} \right] = \delta_{\mathbf{r},\mathbf{s}} \left( \delta_{n,m} - c^0_n \, c^0_m \right),
\end{equation}
the calculation of which is made explicit in App.~\ref{app_commutation_relations}. The correction term on the right-hand side of Eq. (\ref{commutations}) serves to remove those unphysical degrees of freedom that are introduced by the local gauge invariance of the Gutzwiller ansatz (\ref{ansatz}), namely the arbitrariness of the phase of the local wavefunctions $c_n(\mathbf{r})$ at each site $\mathbf{r}$. Result~(\ref{commutations})
generalises to a strongly-interacting Bose system what it is known from the Bogoliubov approach in the homogeneous weakly-interacting Bose gas~\cite{castin}.

Even though in this paper we focus only on Gaussian fluctuations, the inclusion of terms beyond second order in $\delta{\hat{c}}_n$ arising from the quartic hopping term in Eq.~\eqref{lagrangian} does not pose any fundamental difficulty. As in standard Bogoliubov theory, higher-order terms describe interactions between collective modes~(e.g., see Section III in~\cite{LL9}).

\subsection{General remarks on the accuracy of the quantum Gutzwiller method}
\begin{figure}[!htbp]
    \centering
    \includegraphics[width=1.0\linewidth]{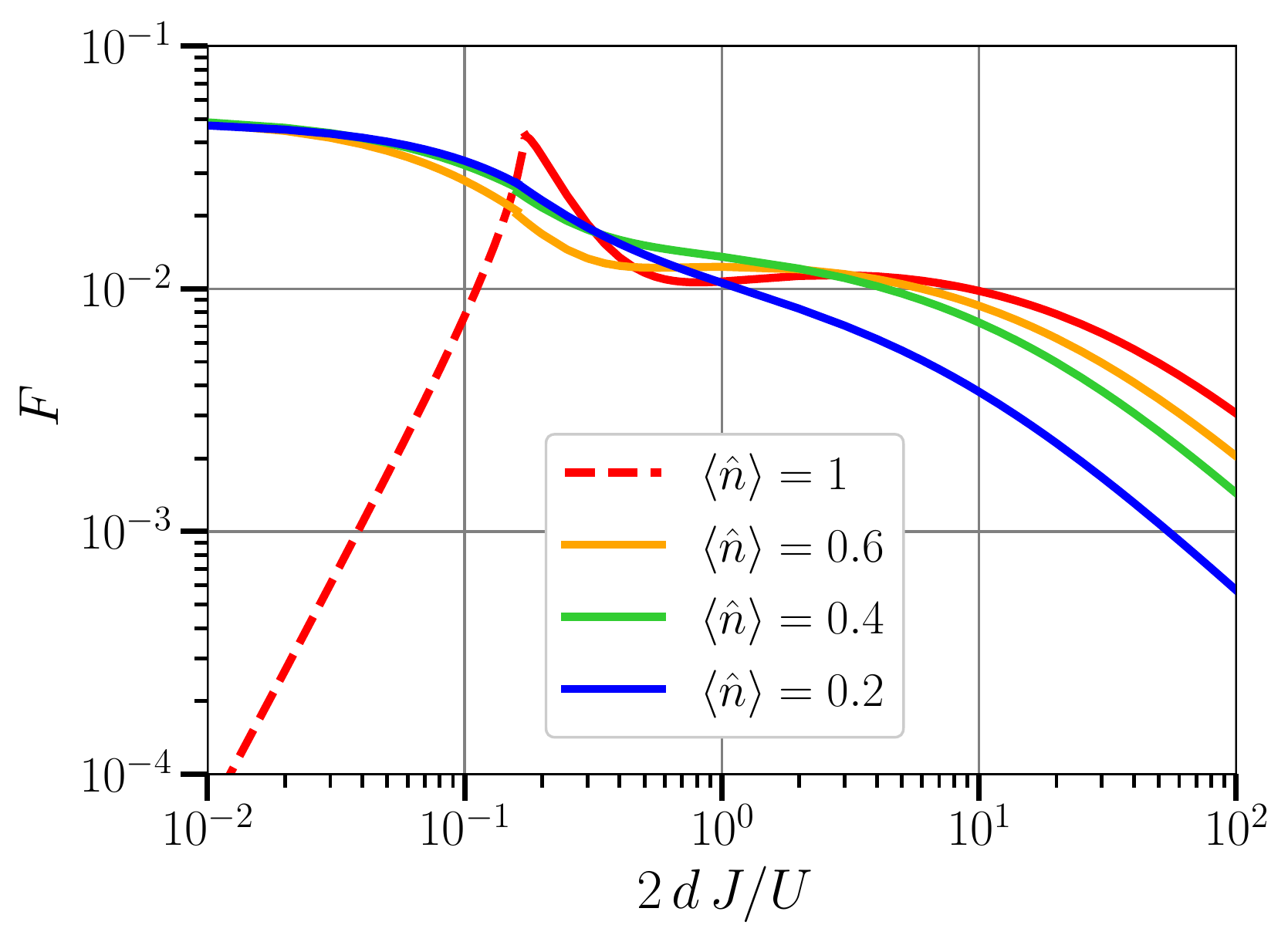}
    \caption{Control parameter $F$ of the theory as defined in (\ref{eq:F}) plotted as a function of $2dJ/U$ for different values of the lattice filling. Dashed and solid lines indicate whether the system is in a Mott insulating (only for $\langle \hat{n} \rangle = 1$) or in a superfluid phase, respectively.}
	\label{fig:figure2}
\end{figure}

The accuracy of the quantum Gutzwiller theory can be quantitatively estimated by looking at the magnitude of the ``quantum fluctuations'' around the Gutzwiller mean-field,
\begin{equation}
    F=1-\langle \hat{A}^2\rangle=
\sum_n \langle \delta c^\dagger_n(\mathbf{r}) \, \delta c_n(\mathbf{r}) \rangle\,,
\label{eq:F}
\end{equation}
which represents the small control parameter of our theory~\footnote{In the standard Bogoliubov theory, the small parameter controlling the accuracy of the Bogoliubov approach has the physical meaning of the non-condensed fraction of the gas. Here it is a mathematical object indicating how much the local wave functions appearing in the Gutzwiller ansatz vary under the effect of quantum fluctuations.}. As it is illustrated in \autoref{fig:figure2}, this quantity remains always very small throughout the phase diagram, suggesting the overall reliability of the quantum Gutzwiller approach: a small value of the quantum fluctuations is in fact a good indication that the nonlinear terms that are not included in (\ref{H_2}) are indeed small and can be neglected.

In particular, for commensurate density, the quantity $F$ approaches zero both in the deep MI (dashed line) and in the deep SF regime, that is in both limits where the Gutzwiller ansatz recovers exactly the ground state of the BH model. As expected, its maximum is located at the transition point.

For non-commensurate densities, $F$ tends to zero in the deep SF regime (where again the Gutzwiller ansatz recovers exactly the ground state) and eventually increases in the strongly interacting superfluid regime for decreasing $J/U\rightarrow 0$. Note that for non-commensurate densities this limit does not correspond to a Mott insulator and the Gutzwiller ansatz is not able to fully capture the ground state.

\subsection{Calculation of the observables}\label{protocol}

In this subsection we summarize the protocol that we use to compute physical observables within the quantum Gutzwiller theory.
The evaluation of the expectation value for any observable $\langle \hat{O}(\hat{a}^{\dag}_\mathbf{r},\hat{a}_\mathbf{r}) \rangle$ consists in applying the following four-step procedure:
\begin{enumerate}
    \item Determine the expression $\mathcal{O}[ c, c^* ] = \big\langle \Psi_G \big| \hat{O} \big| \Psi_G \big\rangle$ in terms of the Gutzwiller parameters $c_n$ and $c_n^*$;
    
    \item Create the operator $\hat{\mathcal{O}}[ \hat{c}, \hat{c}^{\dagger}]$ by replacing the Gutzwiller parameters in $\mathcal{O}\left[ c, c^* \right]$ by the corresponding operators $\hat{c}_n{\left( \mathbf{r} \right)}$ and $\hat{c}^{\dagger}_n{\left( \mathbf{r} \right)}$ without modifying their ordering;
    
    \item  Expand the operator $\hat{\mathcal{O}}$ order by order in the fluctuations $\delta \hat{c}_n$ and $\delta \hat{c}^{\dag}_n$, taking into account the dependence of the operator  $\hat{A}$ on the fluctuation operators. The contribution of $\hat{A}$ may be of fundamental importance when higher orders in the fluctuations become relevant;
    
    \item For the specific case considered in this work of negligible interactions between excitation modes, invoke Wick theorem to compute the expectation value of products of operators on Gaussian states -- such as ground or thermal states obtained from $H^{(2)}$.
\end{enumerate}







In the following, we apply this protocol to compute $\langle \hat{ \cal{O}}\rangle$, where the expectation value is intended to be evaluated on the Bogoliubov vacuum, i.e. ${\hat b}_{\mathbf k,\alpha}|0\rangle=0$.

\subsection{Putting the method into perspective}
\label{subsec:perspective}
Before proceeding with the presentation of the predictions of our theory, it is worth  shortly commenting on the relation of our theory with other competing methods.

Our approach owes much to the time-dependent Gutzwiller method in~\cite{krutitsky_navez} where the $c_n(\mathbf{r})$ parameters are considered as $\mathbb{C}$-numbers and not as operators. In the same way as the linearised Gross-Pitaevskii equation can be used together with linear response theory to obtain information on the quantum fluctuations~\cite{Stringari2018}, the time dependent Gutzwiller approach would give the same results as our method for a number of properties where only quadratic fluctuations are important.

When only Gaussian fluctuations above the MF result are considered, our approach to the BH model has strong similarities to including quantum fluctuations by slave boson techniques, as done, e.g., in the comprehensive work by Fr\'erot and Roscilde \cite{roscilde}. One important difference from this work is however the way in which the observables are calculated: in particular, we never rely on the microscopic reconstruction of the original Bose fields $\hat{a}_{\mathbf r}$ through the operators $\hat{c}_n$'s and, from the very beginning, the dynamical variables of our approach are $\delta \hat c_n$ and $\delta \hat c^\dagger_n$.
It is also worth mentioning that to our knowledge there are no slave boson calculations of the role of quantum fluctuations on the one-body correlation function and on the superfluid density. Even though for such quantities we expect the slave-boson and our approach to give the same results, our method is technically easier and more transparent. Finally, the slave-boson approach has been recently shown to properly interpolate between strong coupling and Bogoliubov approaches in calculating the entanglement entropy~\cite{roscilde}, a property accessible to our approach, but not to the time-dependent Gutzwiller method.

In the next Section, we will show how the quantum Gutzwiller method can reproduce both local and non-local correlations with very high accuracy and successfully compares to Quantum Monte Carlo (QMC) calculations. Moreover, the study of time-dependent problems appears to be a straightforward generalization of the quantum Gutzwiller approach. This is a crucial feature compared to QMC, which can hardly describe dynamical properties.

Dynamical properties can be instead attacked by means of the bosonic version of Dynamical Mean Field Theories (B-DMFT)~\cite{bdmft_vollhardt, bdmft_tong, bdmft_pollet}: while this theory is very accurate for the study of local quantities, it is however poorly reliable for non-local quantities. In addition to not providing physical intuition, both QMC and B-DMFT are computationally much more demanding than the present quantum Gutzwiller approach.

\section{Correlation functions across the MI-SF transition}\label{section_results}

After having introduced the quantum Gutzwiller theory, in the present Section we apply it to the calculation of some relevant correlation functions: (i) the coherence function; (ii) the current-current correlation function and superfluid density; (iii) the density-density correlation function. We compare our results with the predictions of Quantum Monte Carlo calculations, when available, finding striking agreement.

\subsection{Coherence function}\label{section_results_cf}

The single-particle correlation function, referred to also as coherence function, is defined as

\begin{equation}\label{g_1_a}
g^{(1)}{\left( \mathbf{r} \right)} = \frac{\langle \hat{a}^{\dagger}_{\mathbf{r}} \, \hat{a}_{\mathbf{0} } \rangle}{\langle \hat{a}^{\dagger}_{\mathbf{0}} \, \hat{a}_{\mathbf{0}} \rangle} \rightarrow \frac{\langle \hat{\psi}^{\dagger}{\left( \mathbf{r} \right)} \, \hat{\psi}{\left( \mathbf{0} \right)} \rangle}{\langle \hat{\psi}^{\dagger}(\mathbf{0}) \, \hat{\psi}(\mathbf{0}) \rangle}\,,
\end{equation}
where the last expression is the result of the quantization protocol outlined in the previous section.

\begin{figure}[!htbp]
    \begin{minipage}[c]{1.0\linewidth}
	    \centering
		\includegraphics[width=1.0\linewidth]{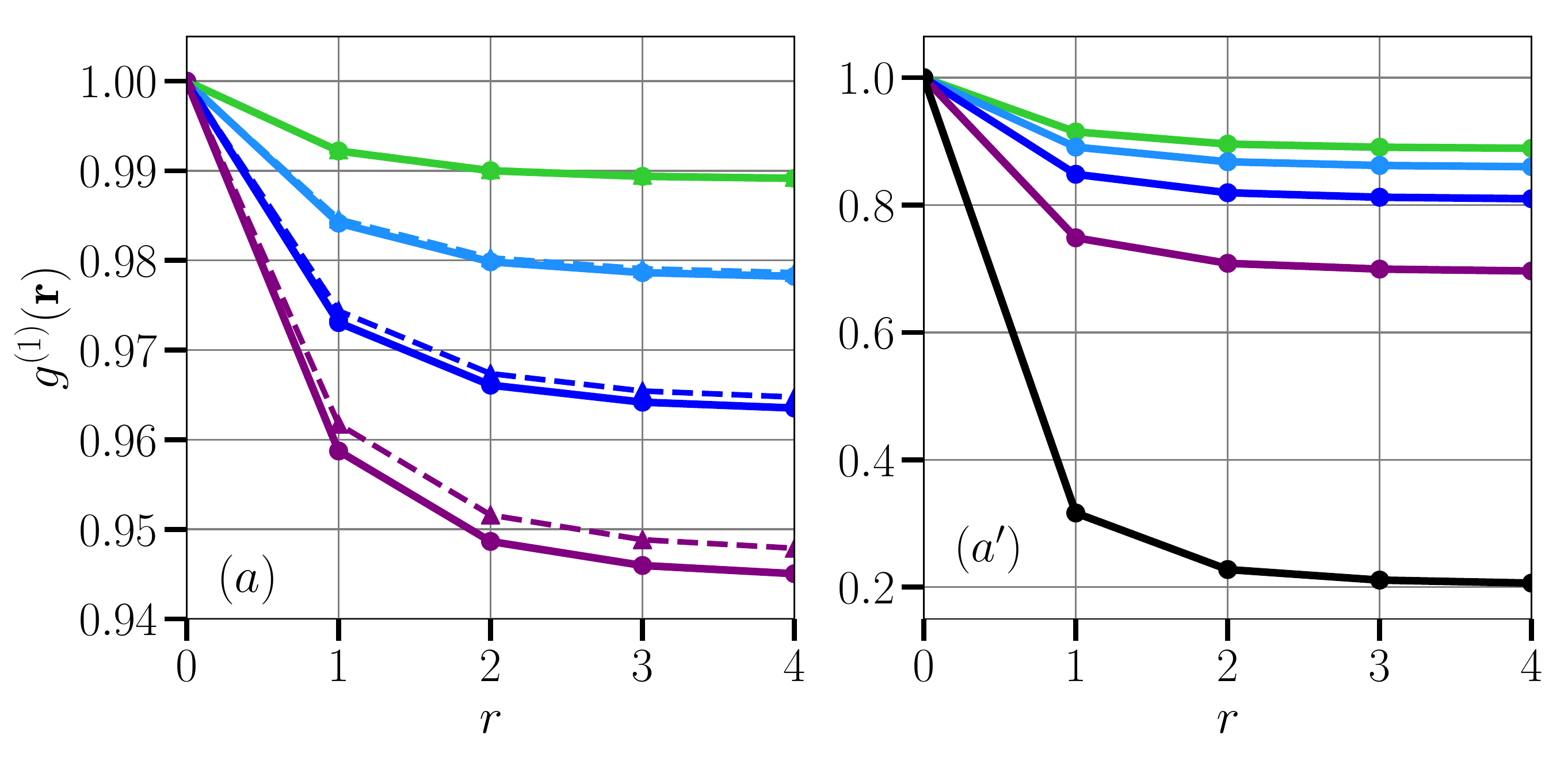}
    \end{minipage}
    \begin{minipage}[c]{1.0\linewidth}
	    \centering
		\includegraphics[width=1.0\linewidth]{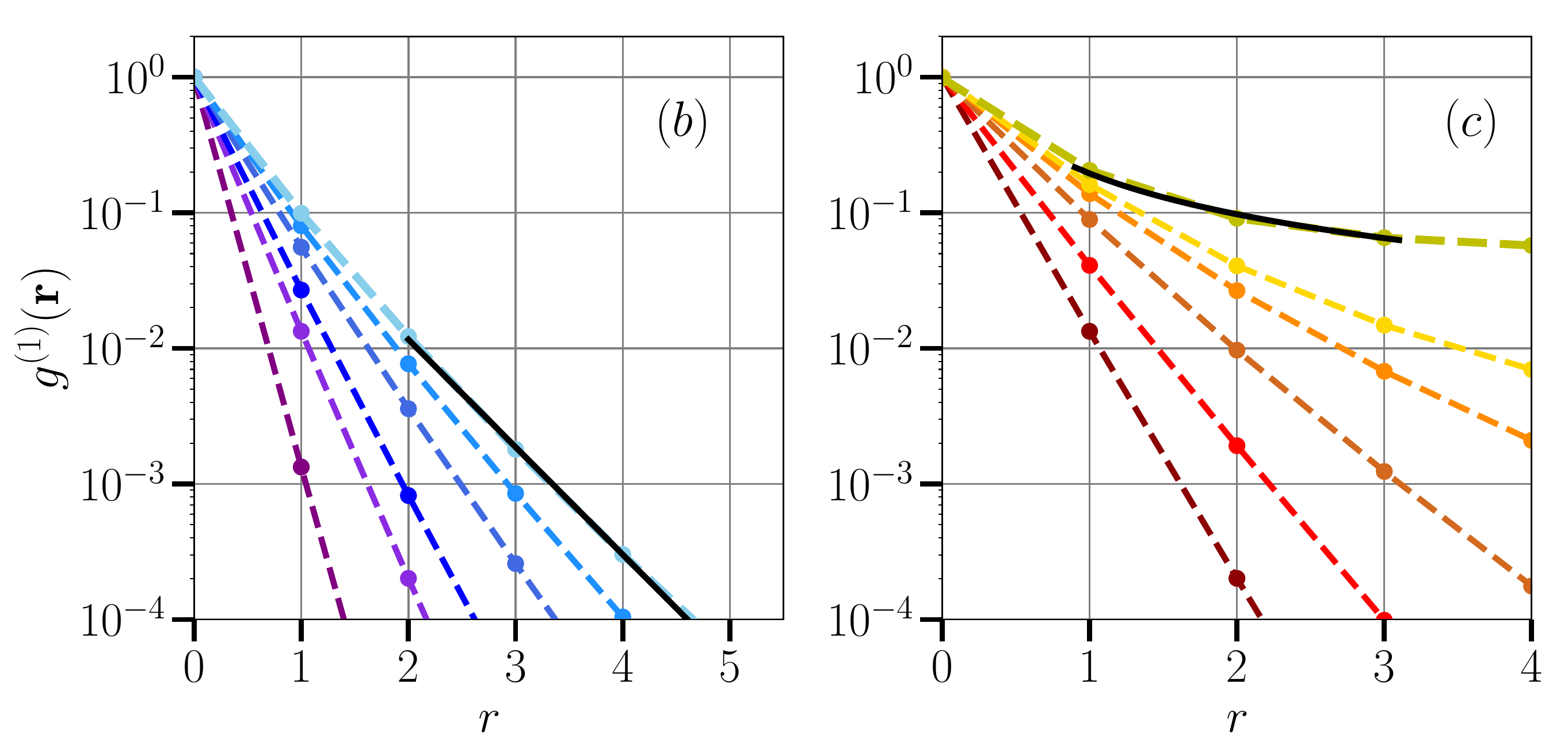}
    \end{minipage}
    \caption{Panel (a): first-order coherence function $g^{(1)}{\left( \mathbf{r} \right)}$ for $\mu/U = \sqrt{2} - 1$ and $2 \, d \, J/U = 2, 3, 5, 10$ going deeper into the SF phase (from bottom to top). Solid and dashed lines refer to the quantum Gutzwiller and Bogoliubov predictions respectively.
    Panel (a'): Gutzwiller predictions of $g^{(1)}{\left( \mathbf{r} \right)}$ for $\mu/U = \sqrt{2} - 1$ and $2 \, d \, J/U = 0.2, 0.4, 0.6, 0.8, 1.0$ approaching the critical point (from top to bottom).
    Panel (b): $g^{(1)}{\left( \mathbf{r} \right)}$ in the MI phase for $\mu/U = 0.2$ and increasing $2 \, d \, J/U = 0.002, 0.02, 0.04, 0.08, 0.11, 0.13, \left( 2 \, d \, J/U \right)_c$ approaching the transition point $\left( 2 \, d \, J/U \right)_c$ from inside the MI lobe (purple to blue lines). Panel (c): $g^{(1)}{\left( \mathbf{r} \right)}$ in the MI phase for $\mu/U = \left( \mu/U \right)_{tip} = \sqrt{2} - 1$ and increasing $2 \, d \, J/U = 0.02, 0.06, 0.12, 0.16, 0.17, \left( 2 \, d \, J/U \right)_{tip}$ towards the tip of the Mott lobe (dark brown to red to gold lines). Dashed lines in panels (b) and (c) are the quantum Gutzwiller predictions, and the (exponential and power-law) fits are displayed as solid black lines.
    }
    \label{fig:figure3}
\end{figure}
Within our protocol the field operator reads
\begin{equation}\label{bose_field_quantization}
\hat{\psi}{\left( \mathbf{r} \right)} = \sum_n \sqrt{n} \, \hat{c}^{\dagger}_{n - 1}{\left( \mathbf{r} \right)} \, \hat{c}_n{\left( \mathbf{r} \right)}. 
\end{equation}
Expanding $\hat{c}_n$ and $\hat{c}^\dagger_n$ to the lowest order in the fluctuations $\delta \hat{c}_n{\left( \mathbf{r} \right)}$, $\delta \hat{c}^\dagger_n{\left( \mathbf{r} \right)}$, one obtains
\begin{equation}\label{bose_field_U_V}
\small
\begin{aligned}
\hat{\psi}{\left( \mathbf{r} \right)} &\approx \sum_n \sqrt{n} \, c^0_{n - 1} \, c^0_n + \sum_n \sqrt{n} \left[ c^0_{n - 1} \, \delta\hat{c}_n{\left( \mathbf{r} \right)} + c^0_n \, \delta\hat{c}^{\dagger}_{n - 1}{\left( \mathbf{r} \right)} \right] \\
&= \psi_0 + \frac{1}{\sqrt{N}} \sum_{\alpha > 0} \sum_{\mathbf{k}} \left( U_{\alpha, \mathbf{k}} \, \hat{b}_{\alpha, \mathbf{k}}e^{i \, \mathbf{k} \cdot \mathbf{r}} + V_{\alpha, \mathbf{k}} \, \hat{b}^{\dagger}_{\alpha, \mathbf{k}}e^{-i \, \mathbf{k} \cdot \mathbf{r}} \right)\,,
\end{aligned}
\end{equation}
where $\psi_0 = \sum_n \sqrt{n} \, c^0_{n - 1} \, c^0_n$ is the order parameter in the ground state and the particle (hole) amplitudes 
\begin{eqnarray}
U_{\alpha, \mathbf{k}}= \sum_{n} \sqrt{n + 1} (c_n^0 u_{\alpha, \mathbf{k}, n+1} +
c_{n+1}^0 v_{\alpha, \mathbf{k}, n}) \\
V_{\alpha, \mathbf{k}}= \sum_{n} \sqrt{n + 1} (c_{n+1}^0 u_{\alpha, \mathbf{k}, n} +
c_{n}^0 v_{\alpha, \mathbf{k}, n+1})
\end{eqnarray}
satisfy the Bogoliubov normalization~\cite{BEC}
\begin{equation}
    \sum_{\alpha} \left( |U_{\alpha, \mathbf{k}}|^2-|V_{\alpha, \mathbf{k}}|^2 \right) = 1.
\end{equation}
In this way, the Bose field \eqref{bose_field_U_V} satisfies the usual canonical commutation relations 
\begin{equation}\label{bose_commutations}
    \left[ \hat{\psi}{\left( \mathbf{r} \right)}, \hat{\psi}^{\dagger}{\left( \mathbf{s} \right)} \right] = \delta_{\mathbf{r}, \mathbf{s}}
\end{equation}
up to second order in the fluctuations.

At the same level of approximation, the normalized zero-temperature one-body coherence function reads
\begin{equation}\label{g_1}
\begin{aligned}
g^{(1)}{\left( \mathbf{r} \right)} \approx
 \frac{|\psi_0|^2 + N^{-1} \sum_{\mathbf{k},\alpha} \left| V_{\alpha, \mathbf{k}} \right|^2 \, \cos{\left( \mathbf{k} \cdot \mathbf{r} \right)}}{|\psi_0|^2 + N^{-1} \sum_{\mathbf{k},\alpha} \left| V_{\alpha, \mathbf{k}} \right|^2}.
\end{aligned}
\end{equation}
In \autoref{fig:figure3}(a-c) we plot the results for $g^{(1)}$ along the different lines at
constant chemical potential in the phase diagram shown in \autoref{fig:figure1}(a). 

In the deep SF phase [panel (a)], the spectral weight is saturated by the Goldstone mode and our prediction for $g^{(1)}(\mathbf{r})$ reduces to the result for the weakly-interacting gas (dashed lines). In the region $2 \, d \, J/U \le 1$ [panel (a')] of strongly-interacting superfluid, the contribution of other excitation modes \cite{menotti_trivedi} starts to become relevant -- and the Bogoliubov approach (not shown) would give much higher asymptotic values. In the MI phase [panels (b, c)], the quantum Gutzwiller method is able to capture an exponentially decreasing coherence $g^{(1)}(\mathbf{r}) \sim \exp{\left( -r/\xi \right)}$ with a finite coherence length $\xi$. A non-vanishing value of $\xi$ provides a first drastic improvement with respect to the mean-field Gutzwiller ansatz, whose factorized form cannot predict any off-site coherence, giving
$
g^{(1)}_{MF}{\left( \mathbf{r} \right)}=(|\psi_0|^2/{n}_0)(1-\delta_{\mathbf{r}, \mathbf{0}})+\delta_{\mathbf{r}, \mathbf{0}}\,,
$
with $n_0 = \sum_n n \, |c_n^0|^2$.

Moreover, the present quantum theory is also able to capture the different critical behaviours of the superfluid-insulator transition depending on whether the transition is approached at integer or non-integer filling.
Approaching the superfluid transition from the Mott phase away from the tip of the Mott lobe, the correlation length $\xi$ of the MI grows but remains bounded [panel (b)]. 
As soon as one enters the SF phase, long-range order suddenly appears as a non-vanishing long-distance coherence, $\langle \hat{a}^{\dagger}_{\mathbf{r}} \, \hat{a}_{\mathbf{0} } \rangle_{ \mathbf{r} \to \infty}\neq 0$: such a quantity physically corresponds to the condensate density $|\psi_0|^2$ and continuously grows from zero as one penetrates the superfluid phase. On the other hand, when approaching the SF phase at the tip of the Mott lobe, the correlation length $\xi$ diverges and a power-law dependence for $g^{(1)}$ is found at the critical point [panel (c)]. 

This remarkable difference is related to the distinct universality class of the MI-SF transition at non-integer or integer filling~\cite{sachdev}. 
In all critical points of the CI transition, either the hole or the particle excitation becomes gapless, respectively the hole excitation below the tip or the particle excitation above the tip. Since the non-trivial short-distance coherence of the MI is due to virtual particle-hole excitations, the exponential decay of $g^{(1)}(\mathbf{r})$ is dominated by the gap of the particle (or hole) excitation which remains finite.
Instead, at the  critical point of the O(2)-transition both the particle and hole modes become gapless (before turning into the Goldstone and Higgs modes on the SF side), which explains the divergent coherence length~\footnote{A more detailed description of the exponential behaviour of $g^{(1)}(\mathbf{r})$ at the CI transition in terms of the properties of the self-energy within the present quantum theory confirming these physical arguments will be the subject of a forthcoming work, F. Caleffi {\sl et al.}, in preparation.}.

It is worth noticing that result \eqref{g_1} can be obtained within the time-dependent Gutzwiller formalism \cite{krutitsky_navez} as the particle response function. This amounts to determine the
time-dependent Gutzwiller wave function resulting from applying as a perturbation the single-particle operator $\hat{a}_{\bf r}$, and identifying $g^{(1)}$ as the linear response function related to the variation of the expectation value of the operator  $\hat{a}_{\bf r}^\dag$ (see the detailed discussion in the Gross-Pitaevskii framework in \cite{Stringari2018}).
However,  the quantum Gutzwiller approach presented in this section provides a simpler, more intuitive and straightforward procedure, not only for the calculation of $g^{(1)}$, but also for many other observables. First of all, the Bogoliubov amplitudes in \eqref{rotation} are calculated once for all and can be used to calculate the expectation value for any combination of operators. 
Then, quantities like the superfluid density, that we are going to compute in the following Sect.~\ref{section_results_sd}, would require very involved calculations using the time-dependent Gutzwiller approach.
Finally, as we are going to show in the next Sect.~\ref{section_results_df}, there are quantities, like the density correlation function $g^{(2)}$, for which the contribution of the normalisation operator $\hat{A}$ is dominant, in particular close and in the Mott phase: while in the time-dependent Gutzwiller approach the inclusion of $\hat{A}$ would be at least a technically cumbersome task, our theory is able to account for the order by order expansion of the normalization operator in a natural and systematic way.

\subsection{Superfluid density}\label{section_results_sd}

The superfluid density $n_s$ is defined through the static limit of the transverse current-current response function~\cite{scalapino_white_zhang_prl, scalapino_white_zhang_prb} of the system, namely
\begin{equation}\label{n_s}
2J{n_s} = \lim_{q_y \to \mathbf{0}} \lim_{\omega \to 0} \Lambda^{x x}_J{\left( q_x = 0, q_y, \omega \right)} - \,\big\langle \hat{K_x} \big\rangle\,,
\end{equation}
where
\begin{equation}
    \hat{K_x}{\left( \mathbf{r} \right)} = -J [ \hat{\psi}^{\dagger}{\left( \mathbf{r} + \mathbf{e}_x \right)} \, \hat{\psi}{\left( \mathbf{r} \right)} + \text{h.c.} ]
\end{equation}
is the local kinetic energy operator along the direction $x$ of the phase twist~\footnote{In the absence of the lattice, due to Galilean invariance the kinetic energy term is replaced by the total density.} and 
\begin{equation}\label{Lambda_def}
\Lambda^{x x}_J{\left( \mathbf{q}, \omega \right)} = -i \int_{0}^{\infty} \!\!\!dt \, e^{i \omega t} \Big\langle \left[ \hat{J}_x{\left( \mathbf{q}, t \right)}, \hat{J}_x{\left( -\mathbf{q}, 0 \right)} \right] \Big\rangle
\end{equation}
is the current-current response function for the current operator 
\begin{equation}
    \hat{J}_x{\left( \mathbf{r} \right)} = i \, J [ \hat{\psi}^{\dagger}{\left( \mathbf{r} + \mathbf{e}_x \right)} \, \hat{\psi}{\left( \mathbf{r} \right)} - \text{h.c.} ].
\end{equation}
The expectation value \eqref{Lambda_def}, as well as the average kinetic energy $\langle \hat{K_x} \rangle$, are calculated applying the protocol outlined in Sect.~\ref{protocol}: the average kinetic energy reads
\begin{equation}\label{K_x}
\langle \hat{K_x} \rangle = -2 \, J \left[ \psi^2_0 + \frac{1}{N} \sum_{\alpha, \mathbf{k}} \left| V_{\alpha, \mathbf{k}} \right|^2 \cos{\left( k_x \right)} \right]\,,
\end{equation}
while the first non vanishing contribution to the response function $\Lambda_J^{xx}$ turns out to be the 4$^{th}$-order correlation
\begin{eqnarray}
\Lambda^{x x}_J({\mathbf{0}, 0 }) = \!\!-4 \, J^2 \sum_{\mathbf{k}, \alpha, \beta} \!\!\frac{\left( U_{\alpha, \mathbf{k}} V_{\beta, \mathbf{k}} - U_{\beta, \mathbf{k}} V_{\alpha, \mathbf{k}} \right)^2}{\omega_{\alpha, \mathbf{k}} + \omega_{\beta, \mathbf{k}}} \sin^2{\left( k_x \right)}. \nonumber \\
\label{Lambda}
\end{eqnarray}
Equation (\ref{Lambda}) reveals the crucial role played by the coupling between different collective modes in suppressing the superfluid stiffness and creating a sort of normal component~\footnote{The very same expression indeed gives the collisionless drag between two Bose gases at zero temperature, where the modes are the in-phase and out-of-phase modes, D. Romito, C. Lobo and A. Recati, arXiv:2002.03955}. 
For the sake of comparison, it is worth reminding that the ground state mean-field Gutzwiller theory would give $\langle \hat{K}_x \rangle = -2 \, J \, \psi^2_0$ and a vanishing current response $\Lambda^{x x}_J({\mathbf{0}, 0 }) = 0$. This leads to equal superfluid and condensate densities, $n_s = |\psi_0|^2$.

\begin{figure}[!htbp]
    \centering
    \includegraphics[width=1.0\linewidth]{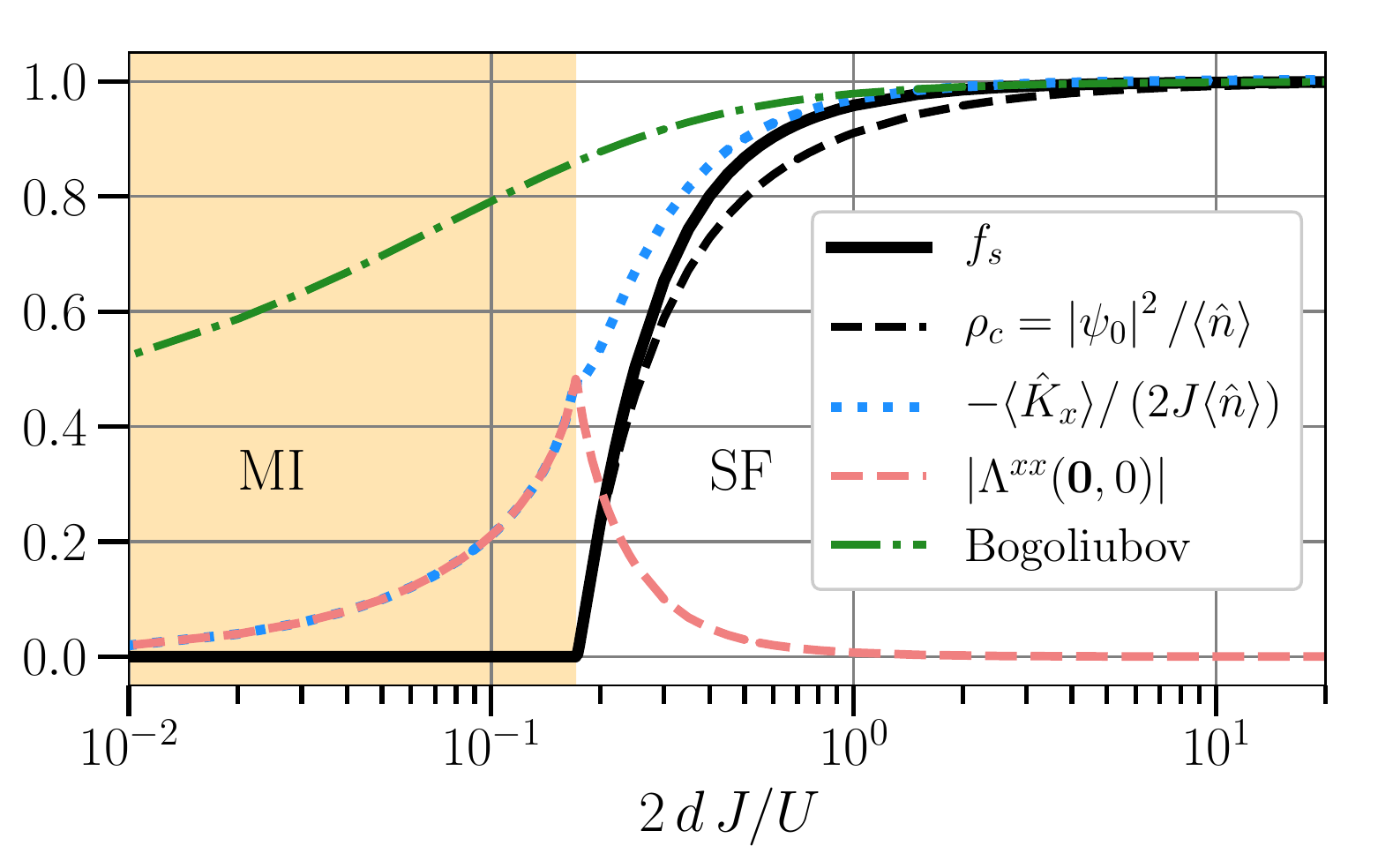}
    \caption{Superfluid fraction $f_s$ along the $\mu/U = \sqrt{2} - 1$ line crossing the tip of the $\langle \hat{n} \rangle = 1$ Mott lobe.  The orange-shaded area indicates the MI region.
    Solid black line: quantum Gutzwiller prediction. Blue dotted and light-pink dashed lines are the contributions to $f_s$ from the kinetic energy $\hat{K_x}$ and the current response $\Lambda_J^{xx}$, respectively. Green dot-dashed line: Bogoliubov approach for the weakly-interacting gas. Black dashed line: condensate fraction $\left| \psi_0 \right|^2/\langle \hat{n} \rangle$. }
	\label{fig:figure4}
\end{figure}

The results of the quantum Gutzwiller method for the superfluid density are illustrated  in \autoref{fig:figure4} by the black thick line for the superfluid fraction $f_s=n_s/{\langle \hat{n} \rangle}$, defined as usual as the ratio of the superfluid density $n_s$ over the total density ${\langle \hat{n} \rangle}$. This quantity tends to unity in the deep SF and approaches zero at the critical point. Throughout the whole SF region, it is always larger than the condensed fraction $|\psi_0|^2$ (black dashed line), defined as usual as the $r \to \infty$ limit of $\langle\hat{a}^\dag_\mathbf{r}\hat{a}_\mathbf{0}\rangle$. In the MI region the superfluid fraction $f_s$ is exactly zero, as expected for a phase that does not exhibit superfluidity.

Further insight is obtained by isolating the two contributions appearing on the right hand side of Eq.~\eqref{n_s}. The current response $\Lambda$ defined in Eq.~\eqref{Lambda} (light-pink dashed line) exhibits a non-monotonic behaviour as a function of $J/U$: it tends to zero in the deep Mott and superfluid phases, while it reaches its maximum at the MI-SF transition point. In the strongly-interacting SF regime, the Goldstone-Higgs vertex almost saturates the sum in the current response \eqref{Lambda} and leads to a complete suppression of $n_s$.
As expected, the kinetic energy Eq.~\eqref{K_x} (see dotted blue line) has as expected a monotonic behaviour, from zero at $J=0$ to the weakly-interacting mean-field value $2J{\langle \hat{n} \rangle}$. In the MI phase, the vanishing $n_s$ results from the perfect cancellation of the short-range virtual particle-hole correlations and the kinetic energy.

For completeness in \autoref{fig:figure4} we also report the the weakly-interacting Bogoliubov prediction~\cite{rey} (green dashed-dotted line), which our result approaches in the limit $2 d J/U \gg 1$. Since it takes into account only the gapless Goldstone mode, such approach leads in particular to a zero current response $\Lambda^{x x}_J({\mathbf{0}, 0 }) = 0$ and thus to an overestimated superfluid density.

\subsection{Density fluctuations}\label{section_results_df}

We consider the normally-ordered equal-time density correlation function 
\begin{equation}
g^{(2)}(\mathbf{r}) = \frac{\langle {\hat a}^\dag_\mathbf{r} {\hat a}^\dag_\mathbf{0} {\hat a}_\mathbf{0} {\hat a}_\mathbf{r} \rangle }{\langle \hat{n}_\mathbf{r} \rangle \, \langle \hat{n}_\mathbf{0} \rangle}.
\end{equation}
\begin{figure}[!htbp]
    \centering
    \includegraphics[width=1.0\linewidth]{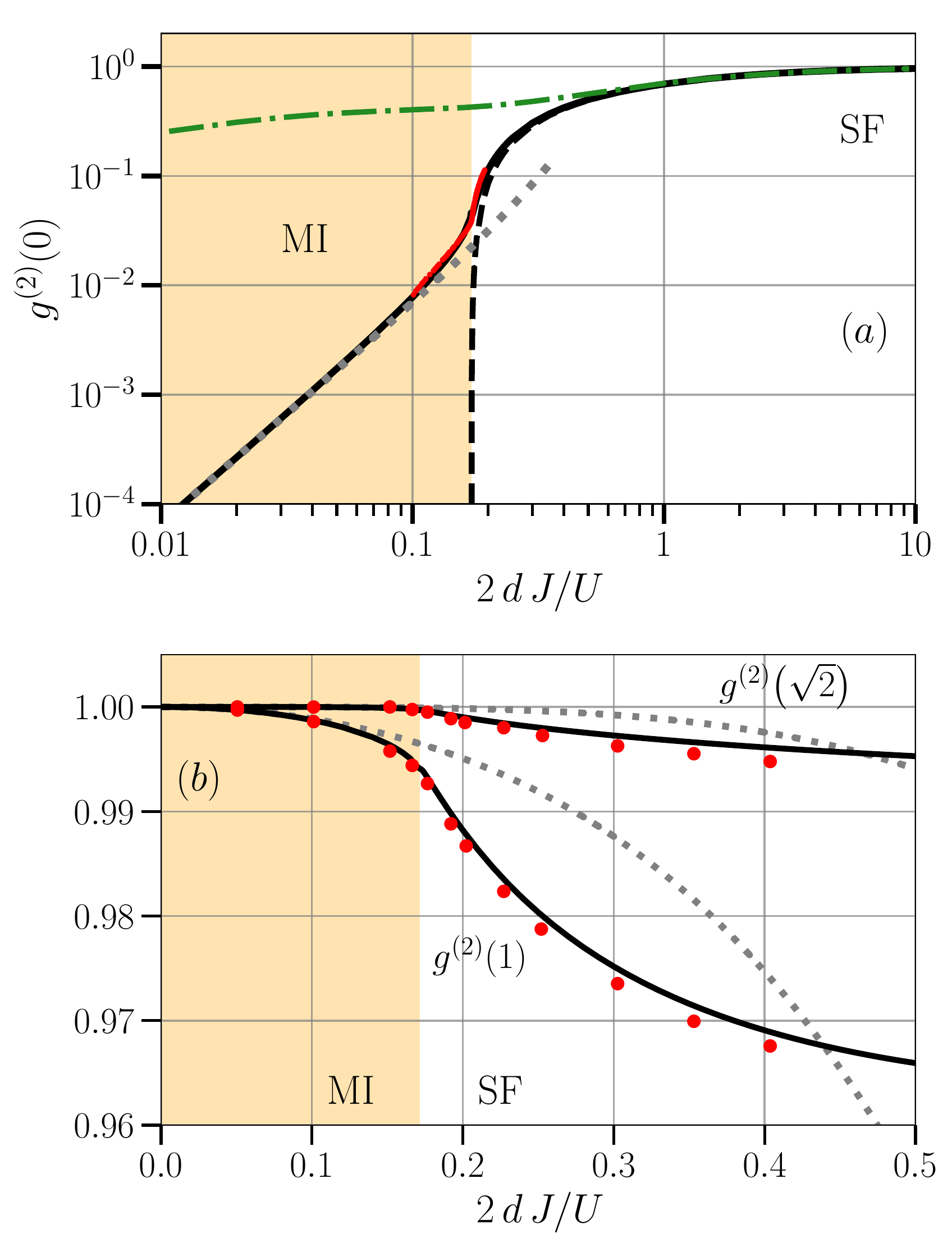}
    \caption{Density-density correlation $g^{(2)}(\mathbf{r})$ as a function of $2dJ/U$ across the $\langle \hat{n} \rangle = 1$ commensurate SF-MI transition. Orange (white) background  identifies the MI (SF) region. Panel (a): On-site correlation function $g^{\left( 2 \right)}(0)$. Black solid line: quantum Gutzwiller method; black dashed line: mean-field Gutzwiller approach; green dot-dashed line: weakly-interacting Bogoliubov theory; red dotted line: Quantum Monte Carlo simulation for a lattice size of $16^3$ sites~\cite{montecarlo}; grey dashed line: strong-coupling perturbation theory~\cite{krutitsky}. Panel (b): Nearest and next-to-nearest density correlations, $g^{\left( 2 \right)}(1)$ and $g^{\left( 2 \right)}(\sqrt{2})$. Black solid  line: quantum Gutzwiller approach; grey dotted line: strong-coupling approximation; red dots: Quantum Monte Carlo calculation for a $5^3$ lattice~\cite{krutitsky}.
    }
    \label{fig:figure5}
\end{figure}
Applying the procedure outlined in Sect.~\ref{protocol}, and since our states are translational invariant, $g^{(2)}(\mathbf{r})$ reads
\begin{equation}\label{g_2_averages}
g^{(2)}(\mathbf{r}) =
\begin{cases}
\left[ \langle \hat{{\cal D}}(\mathbf{0}) \rangle - \langle \hat{{\cal N}}{\left( \mathbf{0} \right)} \rangle \right]\big/\langle \hat{{\cal N}}{\left( \mathbf{0} \right)} \rangle^2, \quad &\mathbf{r} = \mathbf{0}, \\
\langle \hat{{\cal N}}{\left( \mathbf{r} \right)} \, \hat{{\cal N}}{\left( \mathbf{0} \right)} \rangle/\langle \hat{{\cal N}}{\left( \mathbf{0} \right)} \rangle^2, \quad &\mathbf{r} \neq \mathbf{0},
\end{cases}
\end{equation}
where the density $\hat{{\cal N}}{\left( \mathbf{r} \right)} $ and the square density  $\hat{{\cal D}}{\left( \mathbf{r} \right)}$ operators are defined as
\begin{eqnarray}
\hat{\mathcal{N}}{\left( \mathbf{r} \right)} &=& \sum_n n \, \hat{c}^{\dagger}_{n}{\left( \mathbf{r} \right)} \, \hat{c}_n{\left( \mathbf{r} \right)}, \\
\hat{{\cal D}}{\left( \mathbf{r} \right)} &=& \sum_n n^2 \, \hat{c}^{\dagger}_n{\left( \mathbf{r} \right)} \, \hat{c}_n{\left( \mathbf{r} \right)}.
\end{eqnarray}
The expectation values in Eqs.~\eqref{g_2_averages} are evaluated by expanding the operators up to second-order in the $\delta \hat{c}$'s:
\begin{eqnarray}
&\langle \hat{{\cal D}}{\left( \mathbf{0} \right)} \rangle = D_0 + \sum_n \left( n^2 - D_0 \right) \langle \delta \hat{c}^{\dagger}_n{\left( \mathbf{0} \right)} \, \delta \hat{c}_n{\left( \mathbf{0} \right)} \rangle \,,  \label{g_2_expansion_1}  \\
&\langle \hat{{\cal N}}{\left( \mathbf{0} \right)} \rangle = n_0 + \sum_n \left( n - n_0 \right) \langle \delta \hat{c}^{\dagger}_n{\left( \mathbf{0} \right)} \, \delta \hat{c}_n{\left( \mathbf{0} \right)} \rangle  \,, \label{g_2_expansion_2} \\
&\langle \hat{{\cal N}}{\left( \mathbf{r} \neq \mathbf{0} \right)} \, \hat{{\cal N}}{\left( \mathbf{0} \right)} \rangle = n^2_0 + \frac{1}{N} \sum_{\alpha, \mathbf{k}} \left| N_{\alpha, \mathbf{k}} \right|^2 \cos{\left( \mathbf{k} \cdot \mathbf{r} \right)}   \label{g_2_expansion_3} \\ 
&+ \sum_{n, m} \left( n - n_0 \right) \left( m - n_0 \right) \, \langle \delta \hat{c}^{\dagger}_n{\left( \mathbf{r} \right)} \, \delta \hat{c}_n{\left( \mathbf{r} \right)} \, \delta \hat{c}^{\dagger}_m{\left( \mathbf{0} \right)} \, \delta \hat{c}_m{\left( \mathbf{0} \right)} \rangle \nonumber,
\end{eqnarray}
where
\begin{equation}
    N_{\alpha, \mathbf{k}}= \sum_n n \, c_n^0 (u_{\alpha, \mathbf{k}, n}+v_{\alpha, \mathbf{k}, n}).
\end{equation}
Specifically, all second order terms in the expansion arise directly from the normalization operator ${\hat A}(\mathbf r)$. 
This corresponds in saying that, in Eqs.~\eqref{g_2_expansion_1} and \eqref{g_2_expansion_2}, the mean-field quantities $D_0 = \sum_n n^2 \left| c^0_n \right|^2$ and $n_0 = \sum_n n \left| c^0_n \right|^2$ are corrected by the depletion of the mean-field solution itself due to quantum fluctuations, leading to a non-trivial description of local quantum correlations in the ground state. 
Similar features are shared by the non-local correlations in Eq.~\eqref{g_2_expansion_3}, where terms up to fourth-order have to be taken into account. In particular, the role of these higher-order fluctuations dominates for strong interactions, since the spectral amplitude $N_{\alpha, \mathbf{k}}$ appearing in the linear expansion of the density operator $\hat{n}{\left( \mathbf{r} \right)}$ vanishes identically in the Mott phase \cite{krutitsky_navez, BEC}.

The quantum Gutzwiller result \eqref{g_2_averages} for the local density correlation $g^{(2)}(\mathbf{r}=0)$ is shown as a solid black line in the panel (a) of \autoref{fig:figure5}.
On the SF side, the antibunching $g^{(2)}(0) < 1$ due to the repulsive on-site interactions well matches the weakly-interacting Bogoliubov prediction~\cite{rey} in the deep SF (green dashed-dotted line) and increases, faster than the Bogoliubov predictions, when moving towards the Mott transition. On the MI side, while the mean-field Gutzwiller theory (black dashed line) predicts a perfect antibunching $g^{(2)}_{MF}(0) \propto D_0 - n_0 = 0$, the quantum Gutzwiller result \eqref{g_2_averages} is able to account for the virtual excitation of doublon-hole pairs. This leads to  $g^{(2)} \propto J^2$ at low $2dJ/U$, in excellent agreement with strongly-interacting perturbative calculations (gray dotted line)~\cite{krutitsky}. Remarkably, close to and across the critical point, the quantum Gutzwiller theory is in very good agreement with low-temperature Quantum Monte Carlo predictions~\cite{montecarlo} (red dots). In order to compare the results of the two different models, the hopping parameter for the QMC data has been rescaled by a factor $J_c/J_c^{QMC}$ so to make the position of the critical point in the two theories coincide. Note that no other semi-analytical theory is available to describe this region close to and across the critical point.

The role of quantum fluctuations and the accuracy of the method can be further explored by looking at the off-site density correlations function for $|\mathbf{r}|=1$ and $\sqrt{2}$.
In panel (b) of \autoref{fig:figure5} we report the quantum Gutzwiller predictions for $g^{(2)}(1)$ and  $g^{(2)}(\sqrt{2})$ along the $\langle \hat{n} \rangle = 1$ filling line across the tip of the Mott lobe. These curves are successfully compared to available Quantum Monte Carlo data (see~\cite{krutitsky} and references therein) and to strong-coupling perturbation theory, which shows that our theory is accurate across the whole phase transition and correctly interpolates between a strongly-interacting Mott insulator phase and the weakly-interacting Bose gas.

\section{Conclusions}\label{section_conclusions}
In this paper, we have introduced a simple and powerful semi-analytical tool to study the many-body physics of quantum interacting particles on a lattice based on a canonical quantization of the fluctuations around the Gutzwiller ground state. The power of the method has been validated on the archetypal case of the Bose-Hubbard model. In spite of the locality of the initial mean-field Gutzwiller ansatz, the quantization procedure proposed in our work is able to accurately capture very non-local physical features such as the superfluid stiffness of the superfluid phase, the different behaviours of the correlation functions at the different critical points, and the spatial structure of the virtual particle-hole pair excitations on top of a Mott insulator. In particular these last predictions are in quantitative agreement with Quantum Monte Carlo results available in the literature. 
In addition to its quantitative accuracy and to its computational simplicity, the quantum Gutzwiller method has the crucial advantage over other approaches of providing a deep physical insight on the equilibrium state and on the quantum dynamics of the system under investigation.

Due to its flexibility and numerical accessibility, our formalism can be straightforwardly extended to treat inhomogeneous configurations, more exotic hopping and interaction terms, and to deal with more complex forms of the initial ansatz, such as the cluster Gutzwiller wavefunction, which include at least some quantum correlations already in the ground state. An important application of our quantized model for the excitations is the investigation of finite temperature and/or time-dependent problems, including non-equilibrium dynamics. Going beyond the quadratic expansion around the mean-field, our method can in fact naturally incorporate additional nonlinear terms describing interactions between quasi-particles~\cite{Caleffi:MSc} so to describe, e.g., their temporal decay into entangled pairs via multi-branch Beliaev decay processes~\cite{recati2019breaking}. Exciting long term perspectives will be to apply our theoretical framework to those driven-dissipative models that can now be realized in photonic systems~\cite{carusotto2013quantum,ma2019dissipatively,carusotto2019}.

\section*{Acknowledgements}
The authors would like to thank S. Sachdev for useful discussions. C. M., I. C. and A. R. acknowledge financial support from the Provincia Autonoma di Trento and from the FET-Open Grant MIR-BOSE (737017) and Quantum Flagship Grant PhoQuS (820392) of the European Union. M.C. acknowledges financial support from MIUR PRIN 2015 (Prot. 2015C5SEJJ001) and SISSA/CNR project “Super- conductivity, Ferroelectricity and Magnetism in bad metals” (Prot. 232/2015).

\renewcommand\appendixname{APPENDIX}
\appendix
\renewcommand{\theequation}{\thesection.\arabic{equation}}
\section{THE NORMALIZATION OPERATOR \texorpdfstring{$\hat{A}{\left( \mathbf{r} \right)}$}{Ar}}\label{app_bogoliubov_shift}
The normalization operator 
\begin{equation}
    \hat{A}{\left( \mathbf{r} \right)} = \left[ 1 - \sum_n \delta \hat{c}^{\dagger}_n{\left( \mathbf{r} \right)} \, \delta \hat{c}_n{\left( \mathbf{r} \right)} \right]^{1/2}
\end{equation}
entering the expansion of the Gutzwiller coordinates $\hat{c}_n{\left( \mathbf{r} \right)} = \hat{A}_{\mathbf{r}} \, c^0_n + \delta \hat{c}_n{\left( \mathbf{r} \right)}$ is introduced in order to satisfy automatically the local constraint
\begin{equation}\label{constraint}
\sum_n \hat{c}^{\dagger}_n{\left( \mathbf{r} \right)} \, \hat{c}_n{\left( \mathbf{r} \right)} = \hat{\mathds{1}}
\end{equation}
which restricts the action of the Gutzwiller operators to the physical subspace and derives directly from the normalization condition for the original complex-valued parameters $c_n{\left( \mathbf{r} \right)}$. Eq. \eqref{constraint} holds under the additional condition $\sum_n \delta\hat{c}^\dag_n{\left( \mathbf{r} \right)} \, c^0_n  = 0$, namely that the ground state eigenvector is orthogonal the fluctuation field. This condition is assumed at the beginning of the derivation of our theory and guaranteed a posteriori by the spectral properties of the pseudo-Hermitian matrix $\hat{\mathcal{L}}_{\mathbf{k}}$, as discussed in App. \ref{app_L_k}. \\
The physical role of $\hat{A}{\left( \mathbf{r} \right)}$
consists in taking into account the feedback of quantum fluctuations onto the Gutzwiller ground state $c^0_n$ in a self-consistent manner. The importance of such renormalization can be immediately observed in the calculation of local observables as the average square density $\langle \hat{D}{\left( \mathbf{0} \right)} \rangle =\langle{\hat n}^2(\mathbf{0})\rangle$
contributing to the expression of the on-site pair correlation function $g_2{\left( \mathbf{0} \right)}$ presented in Section \ref{section_results_df}. The explicit calculation of $\langle \hat{D}{\left( \mathbf{0} \right)} \rangle$ reads
\begin{eqnarray}\label{D_2_expansion}
&&\langle \hat{D}{\left( \mathbf{0} \right)} \rangle = \sum_n n^2 \langle \hat{c}^{\dagger}_n{\left( \mathbf{0} \right)} \, \hat{c}_n{\left( \mathbf{0} \right)} \rangle  \\
&&= \sum_n n^2 \Big\langle \left[ \hat{A}_{\mathbf{0}} \left( c^0_n \right)^* + \delta \hat{c}^{\dagger}_n{\left( \mathbf{0} \right)} \right] \left[ \hat{A}_{\mathbf{0}} \, c^0_n + \delta \hat{c}_n{\left( \mathbf{0} \right)} \right] \Big\rangle \nonumber \\
&&= \sum_n n^2 \left[ \big\langle \hat{A}^2_{\mathbf{0}} \big\rangle \left| c^0_n \right|^2 + \langle \delta \hat{c}^{\dagger}_n{\left( \mathbf{0} \right)} \, \delta \hat{c}_n{\left( \mathbf{0} \right)} \rangle \right] \nonumber \\
&&= D_0 + \sum_n \left( n^2 - D_0 \right) \langle \delta \hat{c}^{\dagger}_n{\left( \mathbf{0} \right)} \, \delta \hat{c}_n{\left( \mathbf{0} \right)} \rangle \nonumber
\end{eqnarray}
which is the expected result of Eq. \eqref{g_2_expansion_1}. The sum in Eq. \eqref{D_2_expansion} is composed by two terms, one given by quantum fluctuations only and the other, proportional to the mean-field quantity $D_0$, deriving exclusively from the normalization operator via the expectation value $\big\langle \hat{A}^2{\left( \mathbf{0} \right)} \big\rangle$. This second contribution is crucial in the successful predictions for density correlations presented in \autoref{fig:figure5}.

\section{PROPERTIES OF \texorpdfstring{$\hat{\mathcal{L}}_{\mathbf{k}}$}{Lk}}\label{app_L_k}
Fixed the maximal local occupation number at $n_{max}$ for computational purposes, the pseudo-Hermitian matrix $\hat{\mathcal{L}}_{\mathbf{k}}$ is a $2 \, n_{max} \times 2 \, n_{max}$-dimensional object of the form
\begin{equation}\label{L_k}
\mathcal{L}_{\mathbf{k}} =
\begin{pmatrix}
\boldsymbol{H}_{\mathbf{k}} & \boldsymbol{K}_{\mathbf{k}} \\
-\boldsymbol{K}_{\mathbf{k}} & -\boldsymbol{H}_{\mathbf{k}}
\end{pmatrix}
\end{equation}
The $n_{max} \times n_{max}$ blocks $\boldsymbol{H}_{\mathbf{k}}$ and $\boldsymbol{K}_{\mathbf{k}}$ are identical to the ones controlling the Gutzwiller dynamical equations at the linear response level \cite{krutitsky_navez}
\begin{equation}\label{H_k}
\begin{aligned}
H^{n m}_{\mathbf{k}} = &-J{\left( \mathbf{0} \right)} \, \psi_0 \left( \sqrt{m} \, \delta_{n + 1, m} + \sqrt{n} \, \delta_{n, m + 1} \right) \\
&+ \left[ \frac{U}{2} \, n \left( n - 1 \right) - \mu \, n - \hbar \, \omega_0 \right] \delta_{n, m} \\
&- J{\left( \mathbf{k} \right)} \big( \sqrt{n + 1} \sqrt{m + 1} \, c^0_{n + 1} \, c^0_{m + 1} \\
&+ \sqrt{n} \sqrt{m} \, c^0_{n - 1} \, c^0_{m - 1} \big)
\end{aligned}
\end{equation}
\begin{equation}\label{K_k}
\begin{aligned}
K^{n m}_{\mathbf{k}} = &-J{\left( \mathbf{k} \right)} \big( \sqrt{n + 1} \sqrt{m} \, c^0_{n + 1} \, c^0_{m - 1} \\
&+ \sqrt{n} \sqrt{m + 1} \, c^0_{n - 1} \, c^0_{m + 1} \big)
\end{aligned}
\end{equation}
where $J{\left( \mathbf{k} \right)} = 2 \, d \, J - \varepsilon{\left( \mathbf{k} \right)}$ with
\begin{equation}\label{e_k}
\varepsilon{\left( \mathbf{k} \right)} = 4 \, J \sum_{i = 1}^{d} \sin^2{\left( \frac{k_i}{2} \right)}
\end{equation}
is the energy of a free particle on a $d$-dimensional lattice. The ground state energy
\begin{equation}\label{omega_0}
\hbar \omega_0 = -4 \, d \, J \, \psi^2_0 + \sum_n \left[ \frac{U}{2} n \left( n - 1 \right) - \mu \, n \right] \left| c^0_n \right|^2
\end{equation}
is set by the classical evolution of the $c^0_n$'s at the mean-field level and,
shifting the diagonal elements of $\hat{\mathcal{L}}_{\mathbf{k}}$, assures a gapless spectrum in the superfluid phase. 

Among the relevant spectral properties of the pseudo-Hermitian matrix $\hat{\mathcal{L}}_{\mathbf{k}}$ of the quadratic form in Eq. \eqref{H_2}, it is worth mentioning that 
$\left( \underline{c}^0, (\underline{c}^0)^* \right)$ is 
a right eigenvector with zero energy for all momenta $\mathbf{k}$. This fact is intimately related to the physical invariance of the Gutzwiller ansatz \eqref{ansatz} under local phase transformations $c_n(\mathbf{r}) \rightarrow c_n(\mathbf{r})\,e^{i\varphi(\mathbf{r})}$, which reflects into the presence of such spurious eigenvector of $\hat{\mathcal{L}}_{\mathbf{k}}$ both in the superfluid and in the insulating phases.

\section{COMMUTATION RELATIONS}\label{app_commutation_relations}
The commutation relations for the Gutzwiller fluctuation operators $\delta \hat{c}_n{\left( \mathbf{r} \right)}$ can be identified a posteriori once the right eigenvectors $\left( \underline{u}_{\alpha, \mathbf{k}}, \underline{v}_{\alpha, \mathbf{k}} \right)$ of $\hat{\mathcal{L}}_{\mathbf{k}}$ and the exact form of the Bogoliubov rotation \eqref{rotation} are determined. Exploiting the fact that the excitation operators $\hat{b}_{\alpha, \mathbf{k}}$, $\hat{b}^{\dagger}_{\alpha, \mathbf{k}}$ satisfy Bose statistics, it follows that 
\begin{eqnarray}
\label{app_commutations}
&& \left[ \delta \hat{c}_n{\left( \mathbf{r} \right)}, \delta \hat{c}^{\dagger}_m{\left( \mathbf{s} \right)} \right]  \\
&&=\frac{1}{N} \sum_{\alpha, \mathbf{k}} e^{i \, \mathbf{k} \cdot \left( \mathbf{r} - \mathbf{s} \right)} \big( u_{\alpha, \mathbf{k}, n} \, u_{\alpha, \mathbf{k}, m} - v_{\alpha, \mathbf{k}, n} \, v_{\alpha, \mathbf{k}, m} \big), \nonumber
\end{eqnarray}
where all the eigenvector components are assumed to be real. A well-known property of pseudo-Hermitian matrices as $\hat{\mathcal{L}}_{\mathbf{k}}$ is the sum rule \cite{castin}
\begin{equation}\label{sum_rule}
\sum_{\alpha} \left( u_{\alpha, \mathbf{k}, n} \, u_{\alpha, \mathbf{k}, m} - v_{\alpha, \mathbf{k}, n} \, v_{\alpha, \mathbf{k}, m} \right) = \delta_{n, m} - c^0_n \, c^0_m
\end{equation}
that follows from the fact that formally the ground state eigenvector $\left( \underline{c}^0, (\underline{c}^0)^* \right)$
is a zero-energy eigenvector of $\mathcal{L}_{\mathbf{k}}$ 
and can be projected out of the spectral decomposition of the quadratic form \eqref{H_2} when considering only states with finite positive energy in the spectral decomposition (see App. \ref{app_L_k}). Inserting the expression \eqref{sum_rule} into Eq. \eqref{app_commutations}, we obtain the expected quasi-bosonic commutation relations \eqref{commutations}.

\section{QUANTUM CORRECTIONS}
\label{app:d2}

\begin{figure}[!htbp]
    \centering
    \includegraphics[width=1.0\linewidth]{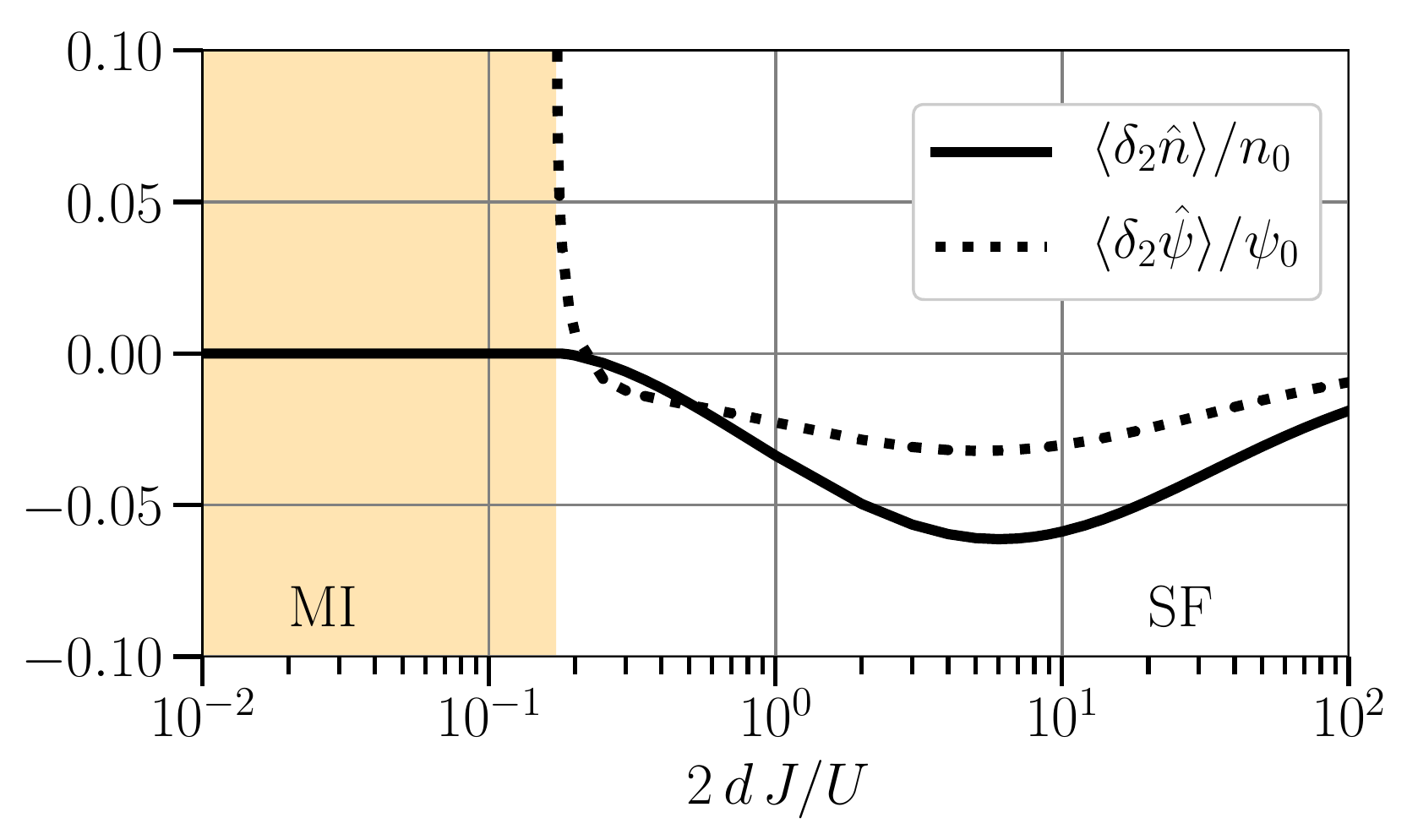}
    \caption{Quantum corrections to the local density (solid line) and order parameter (dotted line) at fixed mean-field density $n_0 = 1$. Orange (white) background indicates the MI (SF) phase.}
    \label{fig:figure6}
\end{figure}

In this Appendix we report explicitly how the order parameter $\psi$ and the the density $n$ are affected by the quantum fluctuations. Both quantities are modified due to the contribution of the normalization operator $\hat{A}$:
\begin{eqnarray}
&&\langle \hat {\mathcal{N}}(\mathbf{r}) \rangle =n_0+\langle \delta_2 \hat n \rangle\nonumber\\
&=& n_0 + \sum_n \left( n - n_0 \right) \langle \delta \hat{c}^{\dagger}_n{\left( \mathbf{r} \right)} \, \delta \hat{c}_n{\left( \mathbf{r} \right)} \rangle\; , \nonumber\\
&&\langle \hat \psi(\mathbf{r}) \rangle =\psi_0+\langle \delta_2 \hat \psi \rangle \nonumber\\
&=&\psi_0 + \sum_n(\sqrt{n} \, \langle \delta \hat{c}^{\dagger}_{n - 1}{\left( \mathbf{r} \right)} \, \delta \hat{c}_n{\left( \mathbf{r} \right)} \rangle -\psi_0  \langle \delta \hat{c}^{\dagger}_n{\left( \mathbf{r} \right)} \, \delta \hat{c}_n{\left( \mathbf{r} \right)} \rangle)\nonumber .
\end{eqnarray}
The corrections at fixed mean-field density $n_0=1$ are reported in \autoref{fig:figure6}. The correction for the density is always very small being zero in the MI phase as well as approaching the non-interacting regime. The order parameter correction is also very small, becoming relevant only extremely close to the transition to the MI phase, where $\psi_0\rightarrow 0$ (notice that on the scale of the Figure the maximum correction is still only $10\%$). Notice also the change in sign of the correction.

Corrections to the mean-field Gutzwiller phase diagram shown in \autoref{fig:figure1}(a) can be obtained by means of a self-consistent calculation including the back-reaction of the quantum fluctuations onto the initial classical Gutzwiller wavefunction. This will be the subject of future work.

\bibliographystyle{apsrev4-1}
\bibliography{apssamp}

\end{document}